\RequirePackage{silence}
\documentclass[pdflatex,sn-mathphys]{sn-jnl}

\usepackage{silence}
\usepackage{amsmath}
\usepackage{bibunits}
\usepackage{booktabs}
\usepackage{cancel}
\usepackage{etoolbox}
\usepackage{graphicx}
\usepackage{lineno}
\usepackage{placeins}
\usepackage{subcaption}

\defaultbibliography{sn-bibliography}
\defaultbibliographystyle{sn-mathphys}

\makeatletter
\newcommand*{\newbibstartnumber}[1]{
  \apptocmd{\thebibliography}{
    \global\c@NAT@ctr #1\relax
    \addtocounter{NAT@ctr}{-1}
  }{}{}
}
\makeatother

\graphicspath{{figures}}

\unnumbered

\begin{document}

\title[Deciphering Earth's Topography Using ML]{Hemispheric Dichotomy of Mantle Dynamics Revealed by Machine Learning} 

\author*[1]{\fnm{Adam J.} \sur{Stewart}}\email{adamjs5@illinois.edu}
\author[2]{\fnm{Yanchong} \sur{Li}}\email{yl79@illinois.edu}
\author[2]{\fnm{Zebin} \sur{Cao}}\email{zebinc2@illinois.edu}
\author*[2]{\fnm{Lijun} \sur{Liu}}\email{ljliu@illinois.edu}

\affil[1]{\orgdiv{Department of Computer Science}, \orgname{University of Illinois at Urbana-Champaign}, \orgaddress{\street{201 North Goodwin Avenue}, \city{Urbana}, \postcode{61801}, \state{IL}, \country{USA}}}
\affil[2]{\orgdiv{Department of Earth Science \& Environmental Change}, \orgname{University of Illinois at Urbana-Champaign}, \orgaddress{\street{1301 West Green Street}, \city{Urbana}, \postcode{61801}, \state{IL}, \country{USA}}}

\abstract{Past efforts to interrogate the mantle's contribution to Earth's topography suffer from inadequate in-situ measurements of true bathymetry and the involvement of an empirical plate model whose presumed lithospheric density profile interferes with the interpretation of other mantle forces. Here,  we introduce a machine learning algorithm that estimates the oceanic residual topography based solely on surface and crustal attributes, providing a more objective proxy for the topographic contribution from the mantle. Ablation studies show that seafloor age is the most important factor, while other properties help further improve the fit to bathymetry. The resulting residual topography has notably smaller amplitudes than previous estimates, indicating that more surface features can be explained by crustal properties than previously thought. This exercise, designed to allow detection of long-wavelength signals, uncovers a more prominent degree-one topographic component than the degree-two pattern, both with much smaller amplitudes than previously thought. We suggest that this result reflects a hemispheric contrast in mantle dynamics, especially the two large low shear-wave velocity provinces (LLSVPs). This inference is further strengthened by quantifying the respective topographic contributions from lithospheric isostasy and dynamic topography. We conclude that mantle convection beneath the Atlantic is more vigorous than that beneath the Pacific, likely due to their different thermal-chemical states.}


\maketitle


Although the operation of plate tectonics on Earth's surface is well accepted, how the underlying mantle functions remains less clear. One important approach to deciphering the style of mantle convection is through examining ocean bathymetry. The systematic deepening of Earth's oceans away from mid-ocean ridges is generally considered isostatic subsidence of the lithosphere as it ages. This process can be described through the plate cooling model (PSM~\cite{parsons1977analysis}, GDH1~\cite{stein1992model}, H13~\cite{hasterok2013heat}), while deviations from this subsidence trend are usually attributed to non-conductive processes beneath the crust, a signal termed \textit{residual} (observation) or \textit{dynamic} (modeling) topography. 

Among the various mantle dynamic processes affecting topography, the existence and amplitude of the degree-2 superswell due to mantle upwelling originated from the two large low shear-wave velocity provinces (LLSVPs) are contentious. Geoid modeling with buoyant LLSVPs favors large-amplitude (1--2~km) dynamic uplift at this wavelength~\cite{hager1985lower,yang2016dynamic}, a prediction challenged by recent in-situ measurement of oceanic residual topography~\cite{hoggard2020global}. However, the latter study was also called into question by its spatially limited data coverage and the adoption of the empirical plate cooling model as a reference level, both of which may prohibit representation of the true convective state of the underlying mantle~\cite{yang2016dynamic,yang2017oceanic}. More recent works~\cite{steinberger2016topography,davies2019earth}, both based on the extrapolated residual topography~\cite{hoggard2016global}, suggest that incorporation of lithospheric density structures in the mantle convection calculation could partially reconcile the above debate, although the respective contributions from the lithosphere and the convective mantle to surface topography still remain inconclusive. Another complication in such exercises is the underestimated density of the continental mantle lithosphere, which may also affect the degree-2 pattern of surface topography~\cite{wang2022topography,wang2022geoid}.

To overcome the sparsity of in-situ measurements and to avoid a priori assumptions about the density structure of oceanic lithosphere, we revisit the problem of oceanic residual topography using a machine learning (ML) algorithm, which effectively covers all ocean basins and utilizes data constraints largely independent of the mantle. The result reveals a power spectrum significantly different from previous studies, with a prominent hemispheric rather than degree-2 pattern of residual topography and another peak around 5--14 degrees. These features outline a different picture of mantle convection and lithosphere dynamics from the traditional view.

\section{Estimating Earth's residual topography}

Our goal is to obtain an objective representation of the global oceanic residual topography with as little mantle contamination as possible. ML using only surface and crustal data could potentially achieve this objective. By training an ML model to associate the measurable quantities with the observed bathymetry, we aim to uncover topographic features that are not determined by shallow crustal properties.

Here, we consider all major structural \textit{features} (seafloor age, layer thickness, seismic velocity, and density) and vertical \textit{layers} (water, ice, sediment, crust, and mantle) of available data that could influence bathymetry. In practice, we utilize constraints for observed topography from ETOPO1~\cite{amante2009etopo1}, seafloor age from multiple recent plate reconstructions~\cite{seton2020global,muller2019global,muller2016ocean,muller2013seawater,muller2008age}, and other structural features and layer properties from CRUST1.0~\cite{laske2013update}. All data are projected onto the same 1\textdegree{} resolution as CRUST1.0. Using the CRUST1.0 crustal type add-on, we filter out data related to continental crust and oceanic plateaus. To prevent the ML model from memorizing the long-wavelength residual topography whose dimension is greater than any individual plate as well as to avoid spatial dependence between data points, we divide the Earth into 7 major tectonic plates: Africa, Antarctica, Australia, Eurasia, North America, South America, and Pacific (Extended Data Fig.~\ref{fig:plate}). The remaining plates are merged into the smallest neighboring plate in order to achieve similarly sized units.

Our training and evaluation procedure uses grouped cross-validation. For each tectonic plate, a model is trained from scratch on the other 6 plates, and then used to make predictions on the 7th plate. This process is repeated for all 7 plates until a prediction is made for every bathymetric data point on Earth. Once predictions have been made for all data points, the predictions are compared to the ground truth from the CRUST1.0 dataset. We compute and report root mean square error (RMSE) and coefficient of determination (R\textsuperscript{2}). Three ML models---ridge regression (RR)~\cite{tikhonov1977solutions}, support vector regression (SVR)~\cite{drucker1996support}, and multi-layer perceptrons (MLP)~\cite{haykin1994neural}---are tested. More descriptions of the data processing and training procedures are provided in Methods and Extended Data Table~\ref{tab:hyper}.

\begin{table}[ht]
    \centering
    \caption{Performance metrics for plate and ML models.}\label{tab:quant}
    \begin{tabular}{lcc}
         \toprule
         Model & RMSE (km) & R\textsuperscript{2} \\
         \midrule
         HS & 1.051 & 0.214 \\
         PSM & 0.689 & 0.662 \\
         GDH1 & \textbf{0.634} & \textbf{0.714} \\
         H13 & 0.652 & 0.697 \\
         \midrule
         RR & 0.592 & 0.751 \\
         SVR & 0.564 & 0.773 \\
         MLP & \textbf{0.476} & \textbf{0.839} \\
         \bottomrule
    \end{tabular}
\end{table}

Regarding the performance of the three ML models in capturing bathymetric information based on the same inputs (Table~\ref{tab:quant}), MLP achieves the lowest RMSE (0.476~km) and highest R\textsuperscript{2} (0.839). To further evaluate the effectiveness of ML models in capturing topographic information, we compare their performance with that of four different physical models, further described in Methods. As expected, the half-space (HS) cooling model~\cite{turcotte2002geodynamics} has the worst performance, with the highest RMSE and lowest R\textsuperscript{2} among all models. Among the three plate models, GDH1 has the best performance. However, all machine learning models we tested have significantly better performance due to their ability to capture information from more crustal layers and features than the plate models. Overall, ML model performance increases with increasingly complex models, with MLP providing the best performance.

\begin{figure}[ht]
    \centering
    \includegraphics[width=0.7\textwidth]{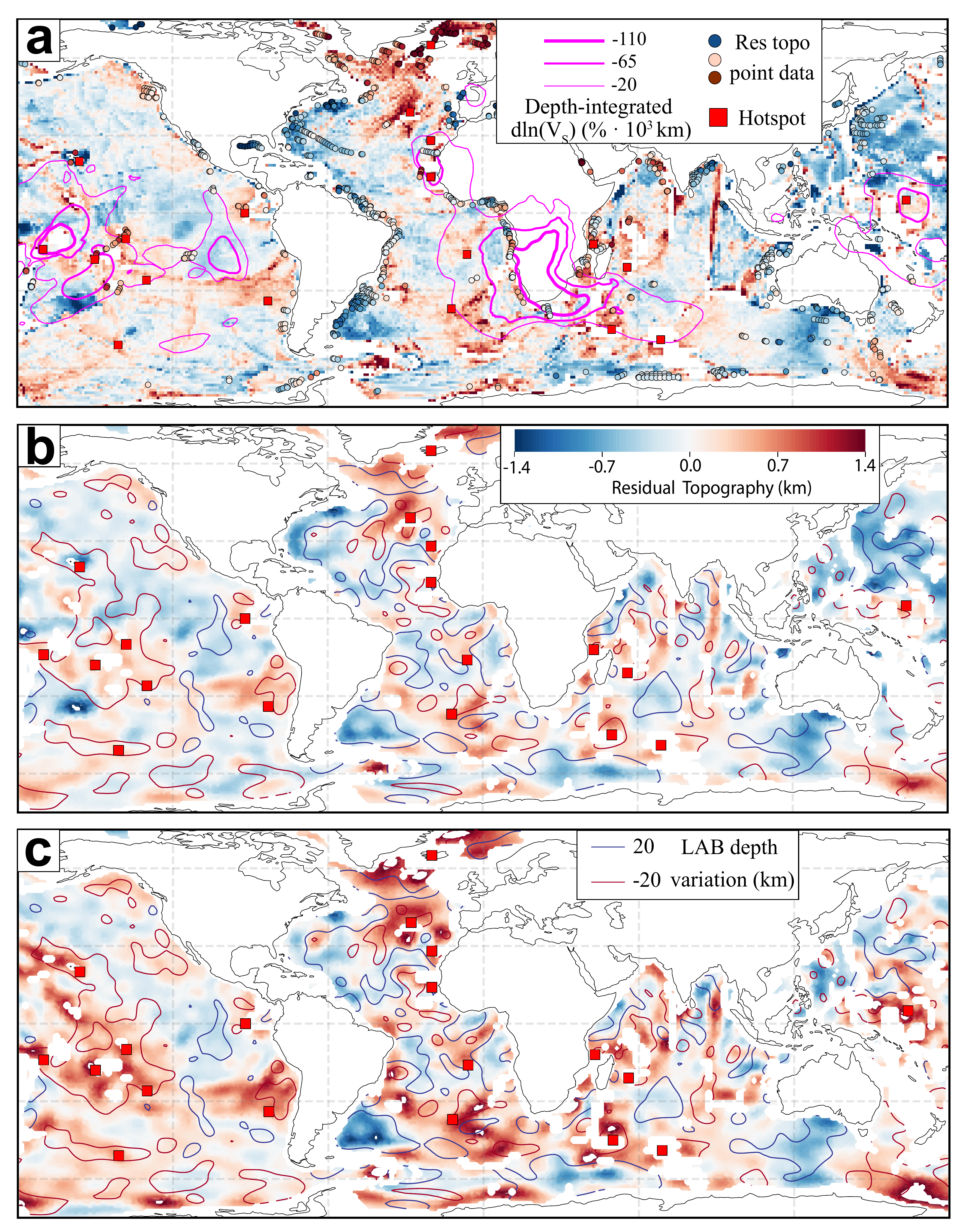}
    \caption{Residual topography derived from a machine learning model and a plate model. a) MLP model of residual topography, overlain with point measurements of oceanic residual topography~\cite{hoggard2016global} and hotspots. Their linear correlation is shown in Fig.~\ref{fig:fig2}a. Purple contours show depth-integrated shear-wave anomalies within the two LLSVPs. b) Low-pass filtered MLP and c) GDH1 residual topography to degree~50. Blue and red contours show larger and smaller LAB depth relative to its global average values.}
    \label{fig:fig1}
\end{figure}

The corresponding residual topography from the best-performing ML model (MLP) and plate model (GDH1) are shown in Fig.~\ref{fig:fig1}. The raw output of the MLP residual topography (Fig.~\ref{fig:fig1}a) contains multi-scale features, with peak topographic values reaching \(\sim\)\,1.5~km. Comparison with the point measurements of residual topography from Hoggard et al.~\cite{hoggard2016global} displays a near 1:1 correlation (Fig.~\ref{fig:fig2}a). This confirms the fidelity of our ML approach in accurately identifying residual topography. The fact that our result covers all ocean basins makes it a more objective representation of global residual topography, as interpolation of point data may introduce arbitrary information to long-wavelength components~\cite{yang2017oceanic}. The observation that nearly all hotspots reside in regions of positive residual topography (Fig.~\ref{fig:fig1}), implying thinner lithosphere or mantle upwelling, further verifies our result.

This residual topography filtered down to spherical harmonic degree~50 (Fig.~\ref{fig:fig1}b) retains all the information of interest for this study while removing local noise. At this scale, the largest amplitude of all topographic signals is below 1~km. This is in contrast to the plate model result which, when filtered to the same wavelengths, displays a notably higher amplitude (\(\sim\)\,1.5~km) (Fig.~\ref{fig:fig1}c). A more striking difference comes from their respective spectrum and correlation with the depth variation of the lithosphere-asthenosphere boundary (LAB), a topic to be further elaborated later (Fig.~\ref{fig:fig2}b,c).
The LAB depth variation is calculated independent of any a priori plate model, where we first construct a global reference LAB depth model by averaging its observed depths for every age, and then map out the local LAB deviation from this reference profile.

\section{Key controlling factors on the results}

In order to determine which features and layers contribute most to model performance, we perform an ablation study. For each feature or layer, we train an MLP model with that feature/layer removed from the dataset to see how it impacts performance. Table~\ref{tab:ablation} lists the performance metrics for each ablation.

\begin{table}[ht]
    \centering
    \caption{Performance metrics for ablation study.}\label{tab:ablation}
    \begin{subtable}[ht]{0.45\textwidth}
        \centering
        \subcaption{Feature ablation.}
        \begin{tabular}{lcc}
            \toprule
            Features & RMSE (km) & R\textsuperscript{2} \\
            \midrule
            No thickness & 0.581 & 0.760 \\
            No $v_p$ & 0.507 & 0.817 \\
            No $v_s$ & 0.489 & 0.830 \\
            No $\rho$ & 0.496 & 0.825 \\
            No age & 0.600 & 0.744 \\
            All features & \textbf{0.476} & \textbf{0.839} \\
            \bottomrule
        \end{tabular}
    \end{subtable}
    \qquad
    \begin{subtable}[ht]{0.45\textwidth}
        \centering
        \subcaption{Layer ablation.}
        \begin{tabular}{lcc}
            \toprule
            Layers & RMSE (km) & R\textsuperscript{2} \\
            \midrule
            No water & 0.483 & 0.834 \\
            No ice & 0.495 & 0.825 \\
            No sediments & 0.625 & 0.722 \\
            No crust & 0.542 & 0.791 \\
            No mantle & 0.574 & 0.766 \\
            All layers & \textbf{0.476} & \textbf{0.839} \\
            \bottomrule
        \end{tabular}
    \end{subtable}
\end{table}

Among all structural features, age turns out to be the most important, with the greatest impact on performance when removed. This may not be surprising given the obvious control of age on bathymetry. One could also expect crustal thickness and density to be important in affecting topography. Indeed, thickness is the second most influential feature. However, density is the second least important. In addition, P- and S-wave velocities contribute a similarly small amount to model performance. The three least important features may correspond to the general lack of lateral crustal variations below oceanic basins.

The layer ablation test reveals that sediments are most impactful to model performance when removed. In fitting seafloor bathymetry, traditional plate models usually ignore this contribution by removing sediments through isostatic unloading. We find that sediments serve as an important factor for ML models to predict bathymetry. In theory, sediments and age are related, with thicker sediment above older seafloors, as could explain their similarly dominant role in model performance. However, they also possess independent information from each other, as can be seen from their spatial mismatch and non-linear correlation (Extended Data Fig.~\ref{fig:seafloor}). This may explain why removing them separately both notably decreases model performance.

Among other layers, crust that contains multiple features (crustal thickness, seismic velocities, and density) is obviously important, but the mantle layer with information on seismic velocity and density also plays a significant role, and is worth further exploration. The water and ice layers containing density and velocity have little impact on model performance. It should be noted that ablation studies do not control for codependent variables. For example, if crustal features are largely a function of age, we can explain why removing the crust from the model does not have as large of an impact on model performance as it would if we had also removed age.

We further test the effect of several published seafloor age datasets. In order to determine how robust our models are against small perturbations in the data, we report the performance of all models trained and evaluated on all these seafloor age datasets in Extended Data Table~\ref{tab:robust}. We find that different seafloor age compilations have minor impacts on model performance. 

\section{Dichotomic residual topography}

\begin{figure}[ht]
    \centering
    \includegraphics[width=0.7\textwidth]{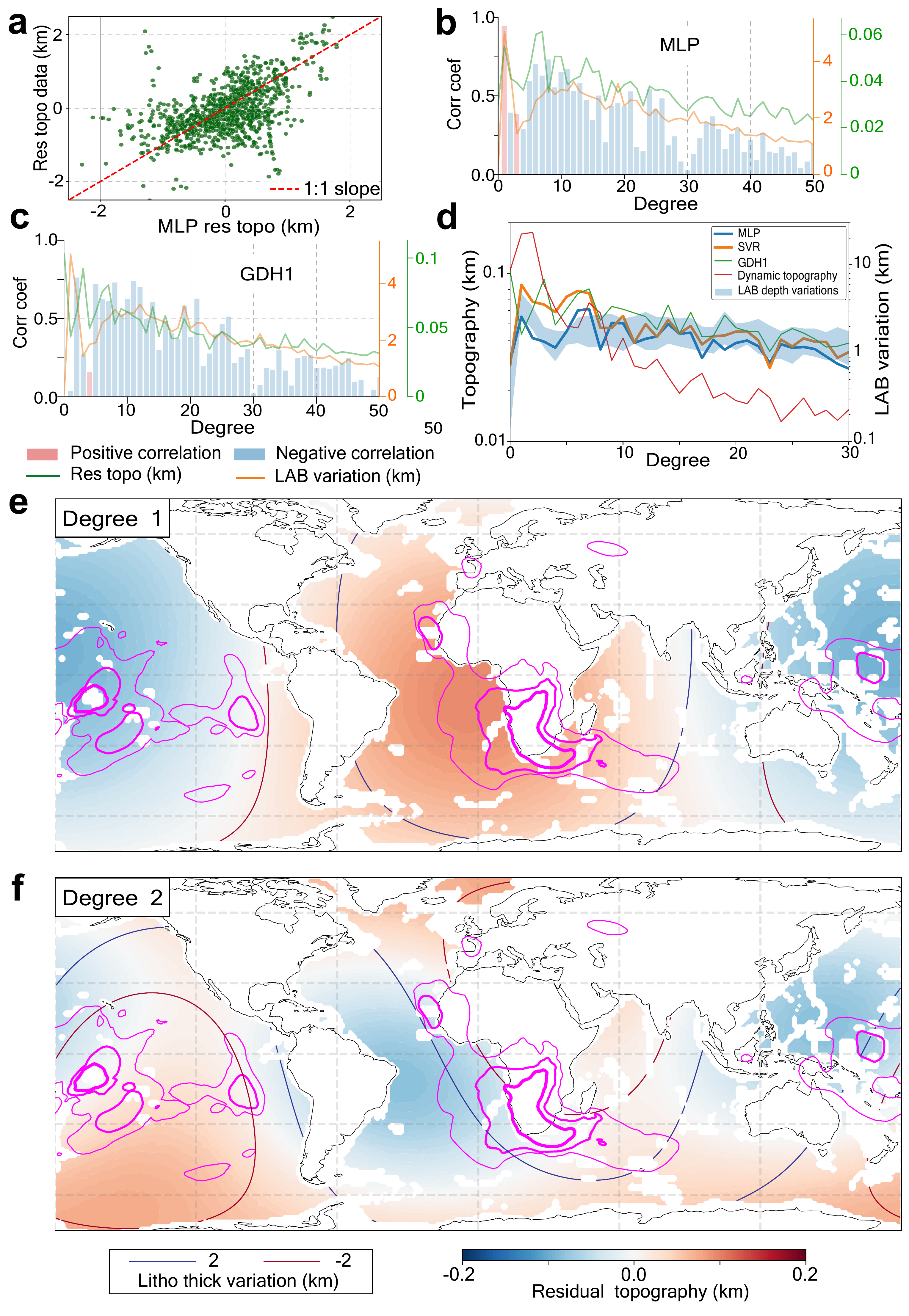}
    \caption{Properties of MLP residual topography. a) Comparison of MLP residual topography with point measurements of oceanic residual topography~\cite{hoggard2016global} shows a linear relationship. b) Correlation of MLP and c) GDH1 residual topography and LAB thickness variation, and their respective spectra expressed in the square root of power with units of km. d) Spectrum of residual (dynamic) topography and LAB thickness variation, all represented in the square root of power with units of km. e) Degree-1 and f) degree-2 components of MLP residual topography.}
    \label{fig:fig2}
\end{figure}

Topographic effects from the convective mantle usually correspond to relatively long wavelengths, which we search for using low-pass filtered residual topography down to degree~30. We observe that the spectra of the ML model results are similar to that of the traditional GDH1 residual topography at degrees higher than 5 (Fig.~\ref{fig:fig2}d), but they contrast sharply at longer wavelengths where they become nearly anti-correlated. This reflects their different capabilities in recognizing long-wavelength topographic features. We suggest that the plate model, assuming age as the only control variable, fails to capture other long-wavelength topographic signals with a shallow origin such as sediment distribution, seismic velocity, and layer thickness.

When evaluated against the tomography-derived dynamic topography (Methods), both the ML and GDH1 spectra show considerably less power at the longest wavelengths (\textless\,degree 4) (Fig.~\ref{fig:fig2}d). However, except at degree 2, the pattern of the power spectrum from the ML results correlates much better with that of dynamic topography than with that of the GDH1 model, an observation lasting up to degree 10 (Fig.~\ref{fig:fig2}d). While this echoes an outstanding issue that the plate-model-based residual topography correlates poorly with dynamic topography at the longest wavelengths~\cite{davies2019earth}, it also demonstrates that ML algorithms may better isolate the topographic  contribution from the convective mantle. 

The most striking difference between the MLP prediction and previous estimates of residual topography (e.g., GDH1) is in their spectra, where the topographic power at very long wavelengths (less than degree 5) is notably less than previously thought (Fig.~\ref{fig:fig2}b--d). Among this wavelength window, the degree-1 component of the ML results stands out most prominently. In map view, this corresponds to a pattern consisting of hemispheric uplift centered on the Atlantic and a hemispheric subsidence centered on the Pacific (Fig.~\ref{fig:fig2}e). This topographic component has an amplitude of \(\sim\)\,150~m.

In stark contrast, the commonly proposed degree-2 residual topography, mostly based on modeled dynamic topography (Fig.~\ref{fig:fig2}d), has negligible power in the ML spectra, especially the MLP prediction. This corresponds to a small amplitude of \textless\,100~m in map view (Fig.~\ref{fig:fig2}f). More surprisingly, the spatial pattern of this degree-2 residual topography is nearly anti-correlated with that of the LLSVPs, with subsidence occurring in the southern Atlantic and west-central Pacific.

\begin{figure}[ht]
    \centering
    \includegraphics[width=0.8\textwidth]{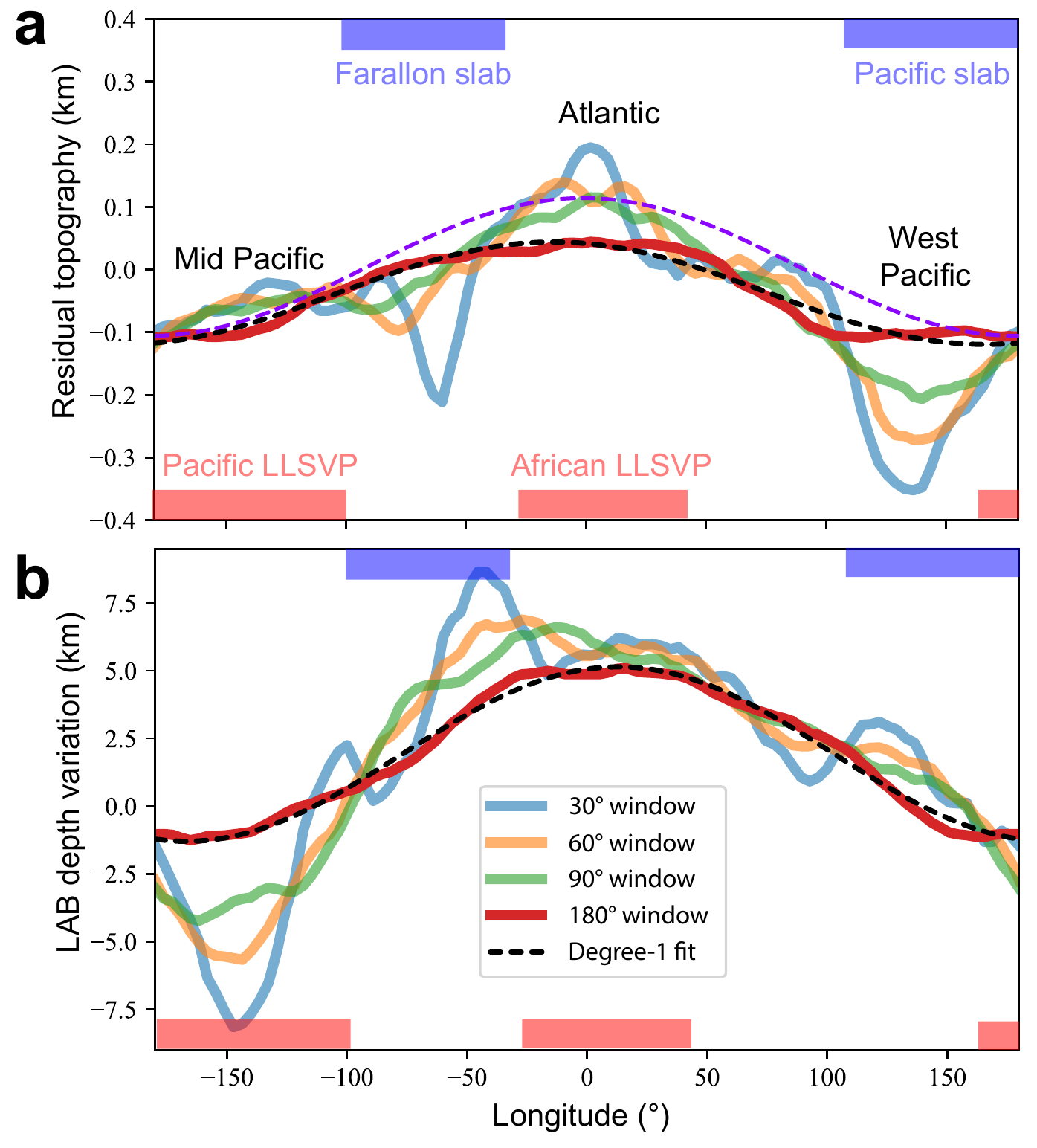}
    \caption{Longitudinal distribution of residual topography and LAB thickness. a) MLP residual topography and b) LAB depth variation averaged over latitude and within longitudinal windows ranging from 30\textdegree{} to 180\textdegree{}. The x-axis shows the center longitude of each window. These stacked profiles progressively approach a degree--1 sinusoidal curve (black dashed line). The purple dashed line is the degree--1 curve for the estimated mantle dynamic topography (Fig.~\ref{fig:fig4}c). Major mantle slabs and LLSVPs are marked.}
    \label{fig:fig3}
\end{figure}

Further comparison of these MLP results with the spectra of LAB thickness variation reveals positive correlation at degree-1 (Fig.~\ref{fig:fig2}b), due to the fact that hemispheric uplift occurs within thick oceanic lithospheres around the Atlantic-African region and subsidence within the Pacific (Fig.~\ref{fig:fig1}b,~\ref{fig:fig2}e). This observation is further confirmed with the longitudinal distribution of residual topography (Fig.~\ref{fig:fig3}a) and LAB depth variation (Fig.~\ref{fig:fig3}b), both of which demonstrate dominant degree-1 power. This result directly contradicts lithospheric isostasy and suggests that the hemispheric residual topography originates from the convective mantle. We note that the same inference cannot be concluded based on the GDH1 result (Fig.~\ref{fig:fig2}c). In contrast, the degree-2 patterns for both MLP and GDH1 results demonstrate a negative correlation with LAB thickness variation (Fig.~\ref{fig:fig2}b,c), permitting but not requiring a possible origin of lithospheric isostasy for this topographic component. Both these findings are different from traditional views.

Another peak within the spectrum of MLP residual topography occurs at degrees 5--14 (Fig.~\ref{fig:fig2}b). Geographically, these signals primarily occur around the LLSVPs and above subducted slabs (Fig.~\ref{fig:fig1}a;~\ref{fig:fig3}), with both prominent uplift and subsidence. From the longitudinal distribution, the uplift signals (relative to the degree-1 background profile) within this wavelength range (corresponding to the 30\textdegree{}--90\textdegree{} averaging windows) also display a hemispheric contrast, with greater amounts of intermediate-wavelength uplift above the African LLSVP than above the Pacific LLSVP (Fig.~\ref{fig:fig3}a). This further supports the different upwelling vigor of the two LLSVPs. The subsidence pattern, however, lacks such a contrast, with clear negative topography above major subduction zones. The slightly less pronounced subsidence above the Farallon slab than above the Pacific slab is consistent with their respective subduction histories, with the former largely finished by now and its slabs sinking mostly into the lower mantle. 

\section{Implications on mantle dynamics}

Further understanding the origin of the MLP residual topography and associated  mantle dynamics requires a quantitative estimate on the respective contributions from the lithosphere and the convective mantle. Here we derive the former from the observed LAB depth variation of oceanic lithosphere via thermal isostasy. Removing the lithospheric topographic contribution from the MLP residual topography generates a plausible dynamic topography due to mantle convection. 

\begin{figure}[ht]
    \centering
    \includegraphics[width=0.85\textwidth]{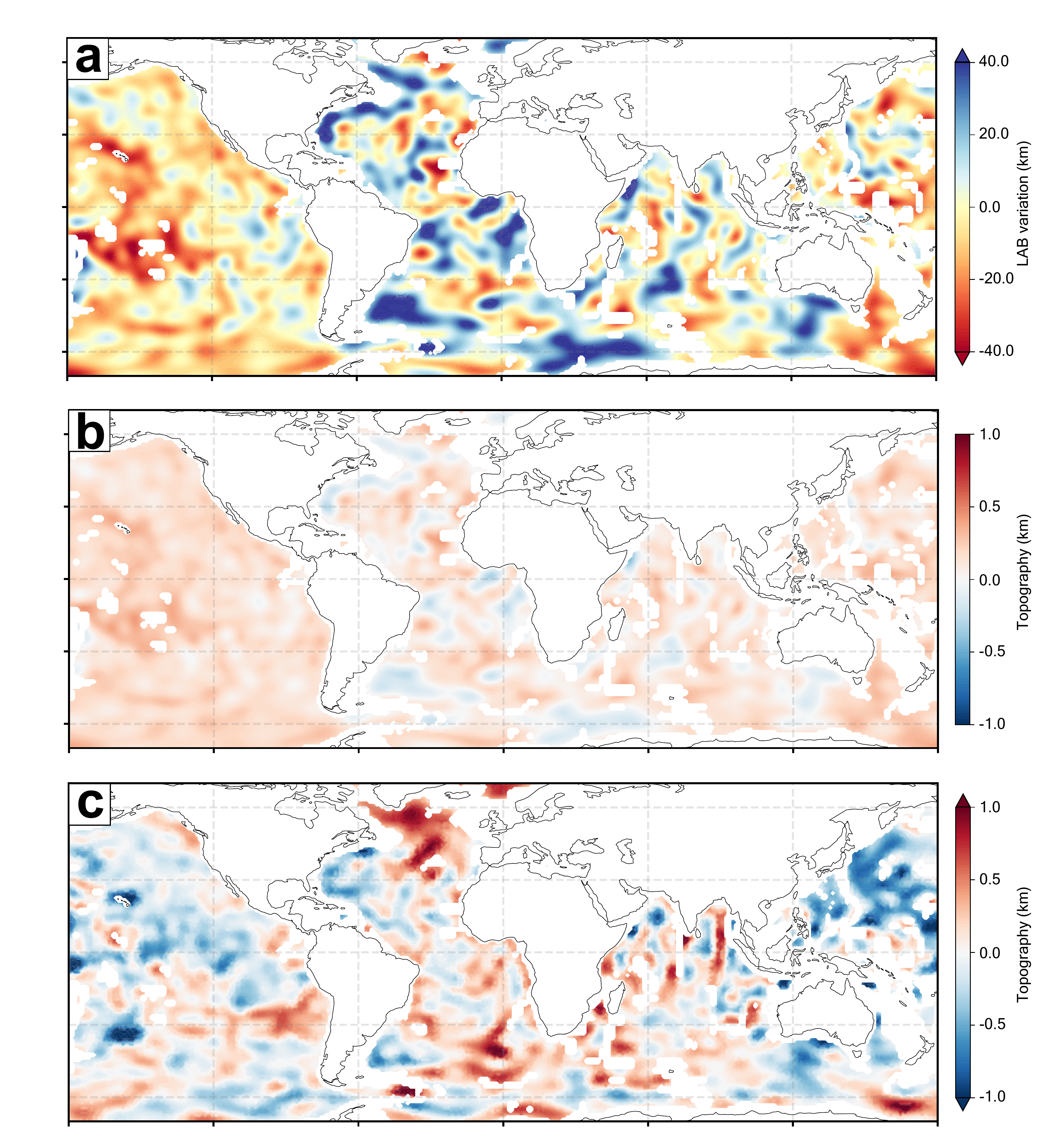}
    \caption{Shallow vs.\ deep origins of residual topography. a) LAB depth variation relative to its global reference profile. b) Estimated contribution to residual topography from the oceanic mantle lithosphere due to its thickness variation. c) Estimated mantle dynamic topography by subtracting b from the original MLP residual topography in Fig.~\ref{fig:fig1}b.}
    \label{fig:fig4}
\end{figure}

The deviation of LAB depth from its global reference profile varies by up to \(\sim\)\,150~km, with overall thicker lithosphere in the Atlantic hemisphere than in the Pacific hemisphere (Fig.~\ref{fig:fig4}a). This observation is confirmed from the longitudinal distribution of LAB depth variation, where a prominent degree-1 pattern exists (Fig.~\ref{fig:fig3}b). We interpret this hemispheric difference in lithospheric thickness as reflecting the corresponding upper mantle temperatures, with the Atlantic side cooler than the Pacific side~\cite{dalton2014geophysical,bao2022relative}. We also observe that the oceanic lithosphere is thicker than the degree-1 baseline above former slabs and thinner above LLSVPs (Fig.~\ref{fig:fig3}b), as is intuitive given the implied temperature effect on lithospheric thickness.

Assuming a depth-averaged temperature perturbation of \(\sim\)\,150~\textdegree{}C near the base of the lithosphere~\cite{turcotte2002geodynamics}, a 150~km thick layer within the lowermost lithosphere generates \(\sim\)\,0.7~km of topographic relief. The pattern of the corresponding isostatic topography variation resembles that of the LAB depth variation, with subsidence above the thick Atlantic lithosphere and uplift above the thin Pacific lithosphere (Fig.~\ref{fig:fig4}b). 

The resulting mantle dynamic topography, computed by removing the above isostatic topography from the MLP residual topography, displays an even more striking degree-1 pattern, as the hemispheric topography contrast is further enhanced (Fig.~\ref{fig:fig4}c), given the positive correlation between the degree-1 mode of MLP residual topography and LAB thickness variation (Fig.~\ref{fig:fig2}b). The peak-to-trough difference in local dynamic topography reaches \textgreater\,2~km, and that of its degree-1 component is nearly twice that of the degree-1 MLP residual topography. The overall pattern of dynamic topography is not affected by the scaling from LAB depth variation to isostatic topography (Extended Data Fig.~\ref{fig:lithscales}). 

The robust pattern of degree-1 dynamic topography reveals that the Pacific mantle generates less uplift than the Atlantic mantle does. We suggest that this reflects that the African LLSVP is more dynamically unstable than the Pacific LLSVP, as may be related to the lower intrinsic density of the former~\cite{yuan2022instability}. The observation that dynamic uplift at both the hemispheric scale and intermediate wavelengths (degrees 5--14) within the Atlantic Ocean supports this inference. The more prominent uplift above the African hemisphere is also consistent with the greater amount of depth-integrated slow seismic anomalies within the African LLSVP relative to the Pacific LLSVP (Fig.~\ref{fig:fig1}a).

This finding, however, is at odds with the reduced upper mantle temperature~\cite{dalton2014geophysical,bao2022relative} and thicker oceanic lithosphere (Fig.~\ref{fig:fig4}a) below most of the Atlantic and Indian Oceans in comparison with the Pacific Ocean. There may be two possible explanations for this. One possibility links to the temporal evolution of the lower  mantle structures, where the African LLSVP is rapidly elevating in response to the converging motion of the surrounding slab piles, many of which (Nazca, Farallon, Pacific, etc.) reached the lowermost mantle during the Cenozoic~\cite{peng2022quantifying}. This tends to generate more long-wavelength surface uplift than the Pacific LLSVP that is farther away from lower mantle slabs. The latter, due to its longer stability, could have accumulated more localized hot materials in the upper mantle to erode the overlying lithosphere, resulting in thinner lithosphere and milder dynamic uplift.

Another possibility concerns the hemispheric compositional difference of the deep mantle. It has been recently proposed that the distinct geochemical signature of the Pacific mantle from the African mantle suggests different influences of the circum-Pacific subduction~\cite{doucet2020distinct}. The African mantle is not only circumvented by multiple slab piles, but also filled with delaminated continental lithosphere beneath much of the Atlantic Ocean~\cite{hu2018modification}. Geodynamic modeling suggests that these continental keels, mostly distributed at upper-mantle depth as seismically observed~\cite{hu2018modification}, are buoyant, thus favoring mantle upwelling and surface uplift~\cite{peng2022fate}. These compositional properties could explain the prominent dynamic uplift and thick oceanic lithosphere in the Atlantic hemisphere, as is largely absent in the Pacific side. 

In summary, our new residual topography inference based on machine learning reveals that the African mantle generates more dynamic uplift than the Pacific mantle. We attribute this hemispheric dichotomy to the contrasting dynamic states of two LLSVPs, where the African LLSVP is still shrinking in response to surrounding subducted slabs, thus generating more uplift than the Pacific side. The residence of cold but buoyant continental lithospheric materials within the African upper mantle prevents lithospheric thinning while maintaining its high topography. Consequently, this study highlights the importance of lateral variation in the thermal-chemical configuration of the deep mantle. 






\section{Methods}\label{sec:methods}

Below, we detail the steps taken to produce our ML results.

\subsection{Data}

The primary dataset used in this work is the CRUST1.0 model~\cite{laske2013update}. This dataset provides measurements of 4 \emph{features} (boundary topography, P-wave velocity, S-wave velocity, and density) across 9 \emph{layers} (water; ice; upper, middle, and lower sediments; upper, middle, and lower crystalline crust; and below the Moho). Observations are recorded on a 1\textdegree{} grid for the entire Earth, with surface topography derived from ETOPO1~\cite{amante2009etopo1}. We also make use of the crustal type add-on during preprocessing, which divides the Earth's crust into 36 categories.

A number of seafloor age datasets produced by Müller et al.~\cite{seton2020global,muller2019global,muller2016ocean,muller2013seawater,muller2008age} are used in this study. The results reported above are produced using the most recent dataset. In order to ensure that our results are robust across all datasets, we also report results for older datasets in Extended Data Table~\ref{tab:robust}.

A plate boundary model produced by Bird~\cite{bird2003updated}---specifically, a version converted into an easy-to-use shapefile~\cite{ahlenius2014world}---is used during cross-validation. We divide the Earth into 7 major tectonic plates: Africa, Antarctica, Australia, Eurasia, North America, South America, and Pacific. The remaining plates are merged into the smallest neighboring plate in order to achieve similarly sized units.

\subsection{Preprocessing}

All datasets are downsampled to the same 1\textdegree{} resolution as CRUST1.0 and spatially joined using GeoPandas~\cite{kelsey_jordahl_2020_3946761}. Using the CRUST1.0 crustal type add-on, we filter out all data except for normal oceanic crust (A1) and young oceanic crust (A0), including all continental crust and oceanic plateaus. The depth to the boundary between lower sediments and upper crystalline crust is chosen as the ground truth that our models attempt to predict. All boundary topographies are converted to layer thicknesses, and the thickness of the water layer is removed from the dataset since this is approximately what we are trying to predict. Finally, all data is standardized by subtracting the mean and dividing by the standard deviation. This standardization is reverted before calculating any performance metrics.

\subsection{Models}

For baseline comparison, we implement the HS~\cite{turcotte2002geodynamics}, PSM~\cite{parsons1977analysis}, GDH1~\cite{stein1992model}, and H13~\cite{hasterok2013heat} models using the hyperparameters reported by the authors. We apply a sediment correction factor, as outlined in the Supplementary Information. We experiment with 3 different machine learning models. All machine learning models are implemented in Scikit-learn~\cite{scikit-learn}.

The simplest model we try is linear regression (LR)~\cite{kenney1962linear} via ordinary least squares. However, our robustness study reveals that outliers in the seafloor age dataset result in wildly different predictions. We instead focus on ridge regression (RR)~\cite{tikhonov1977solutions}, a variant of LR that adds a regularization term to avoid overfitting.

Support vector regression (SVR)~\cite{drucker1996support}, a variant of support vector machines (SVM)~\cite{cortes1995support} designed for regression, is chosen for a more advanced model. Support vector models use a similar loss function as linear regression, but ignore points within a small margin around the decision boundary or prediction. They also allow non-linear solutions.

The most advanced model we experiment with is a multi-layer perceptron (MLP)~\cite{haykin1994neural}, a type of neural network. MLPs are a series of matrix multiplications interspersed with non-linear activation functions. They allow the model to fit arbitrary non-linear functions, but require more training data than other models.

\subsection{Training and evaluation}

Our training and evaluation procedure uses grouped cross-validation. All preprocessed data is divided into 7 major tectonic plates as described above. For each tectonic plate, a model is trained from scratch on the other 6 plates, and then used to make predictions on the 7th plate. This process is repeated for all 7 plates until a prediction is made for every data point on Earth. This plate tectonic-based grouped cross-validation procedure is designed to avoid geospatial dependence between neighboring points seen in prior work~\cite{stewart2020understanding}.

Once predictions have been made for all data points, the predictions are compared to the ground truth from the CRUST1.0 dataset. We compute and report root mean square error (RMSE) and coefficient of determination (R\textsuperscript{2}).

\subsection{Dynamic topography}

We define dynamic topography as the topographic contribution from the convective mantle beneath the lithosphere. We estimate the dynamic topography by solving for the instantaneous global mantle flow using the spherical finite-element code CitcomS~\cite{zhong2008benchmark}. The mantle density structure is converted from a shear wave tomography S40RTS~\cite{ritsema2011s40rts} using an empirical seismic-to-density conversion following a recent study~\cite{wang2022topography}. The resulting density anomalies are further converted to effective temperature perturbations, whose dynamic properties are solved for in a thermomechanical model. 

In practice, we assume an incompressible mantle that satisfies the Boussinesq approximation. The 3D structure of the effective mantle viscosity is dependent both on depth and on temperature~\cite{wang2022topography}. The convective mantle is defined as that below the LAB, estimated as the base of the high shear-wave heterogeneity. In practice, we take 300~km as the LAB depth below continents and 200~km below oceans for the dynamic topography model, similar to previous studies~\cite{davies2019earth,wang2022topography}. The surface dynamic topography is obtained from the sub-lithospheric normal stress.

\backmatter

\bmhead{Data availability}

The CRUST1.0 model used in our experiments is available at \url{https://igppweb.ucsd.edu/~gabi/crust1.html}. All seafloor age datasets can be found on EarthByte: \url{https://www.earthbyte.org/category/resources/data-models/seafloor-age/}. The plate tectonic boundaries shapefile can be found on GitHub: \url{https://github.com/fraxen/tectonicplates}. All predictions made by our ML models are released under a CC-BY 4.0 license and can be found on Hugging Face: \url{https://huggingface.co/datasets/torchgeo/bathymetry}.

\bmhead{Code availability}

In order to ensure the reproducibility of the experiments presented in this paper, all code used is released under an MIT license and can be found on GitHub: \url{https://github.com/adamjstewart/bathymetry}. 




\bmhead{Acknowledgments}

The authors acknowledge the Texas Advanced Computing Center (TACC) at The University of Texas at Austin for providing HPC resources that have contributed to the research results reported within this paper. URL: \url{https://www.tacc.utexas.edu}

This work made use of the Illinois Campus Cluster, a computing resource that is operated by the Illinois Campus Cluster Program (ICCP) in conjunction with the National Center for Supercomputing Applications (NCSA) and which is supported by funds from the University of Illinois at Urbana-Champaign.












\bibliography{sn-bibliography}

\section{Extended data}

\renewcommand\figurename{Extended Data Fig.}
\setcounter{figure}{0}

\renewcommand\tablename{Extended Data Table}
\setcounter{table}{0}

\begin{figure}[htbp]
    \centering
    \includegraphics[width=0.95\textwidth]{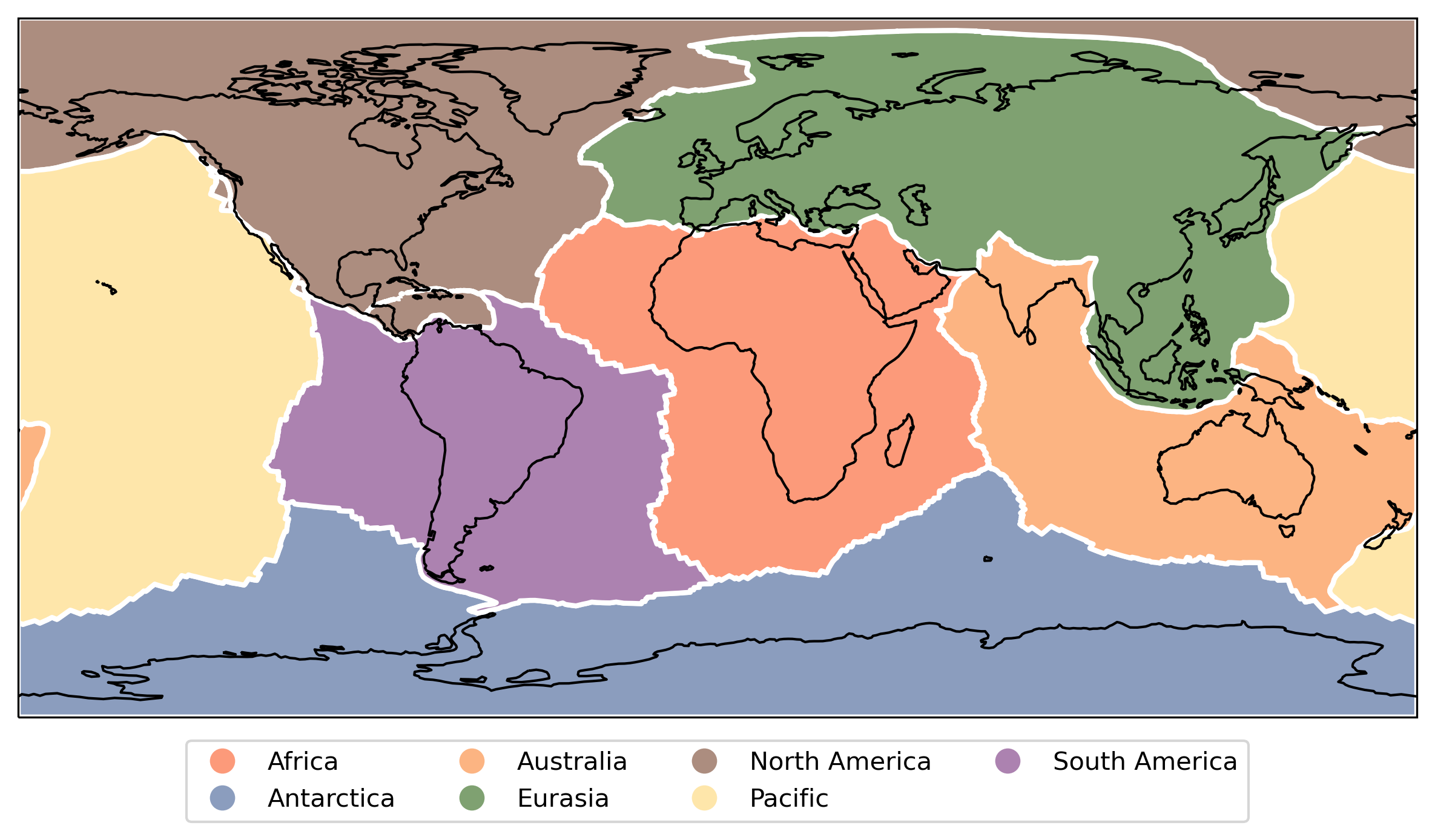}
    \caption{Plate boundaries used in our study.}
    \label{fig:plate}
\end{figure}

\FloatBarrier

\begin{figure}[htbp]
    \centering
    \includegraphics[width=\textwidth]{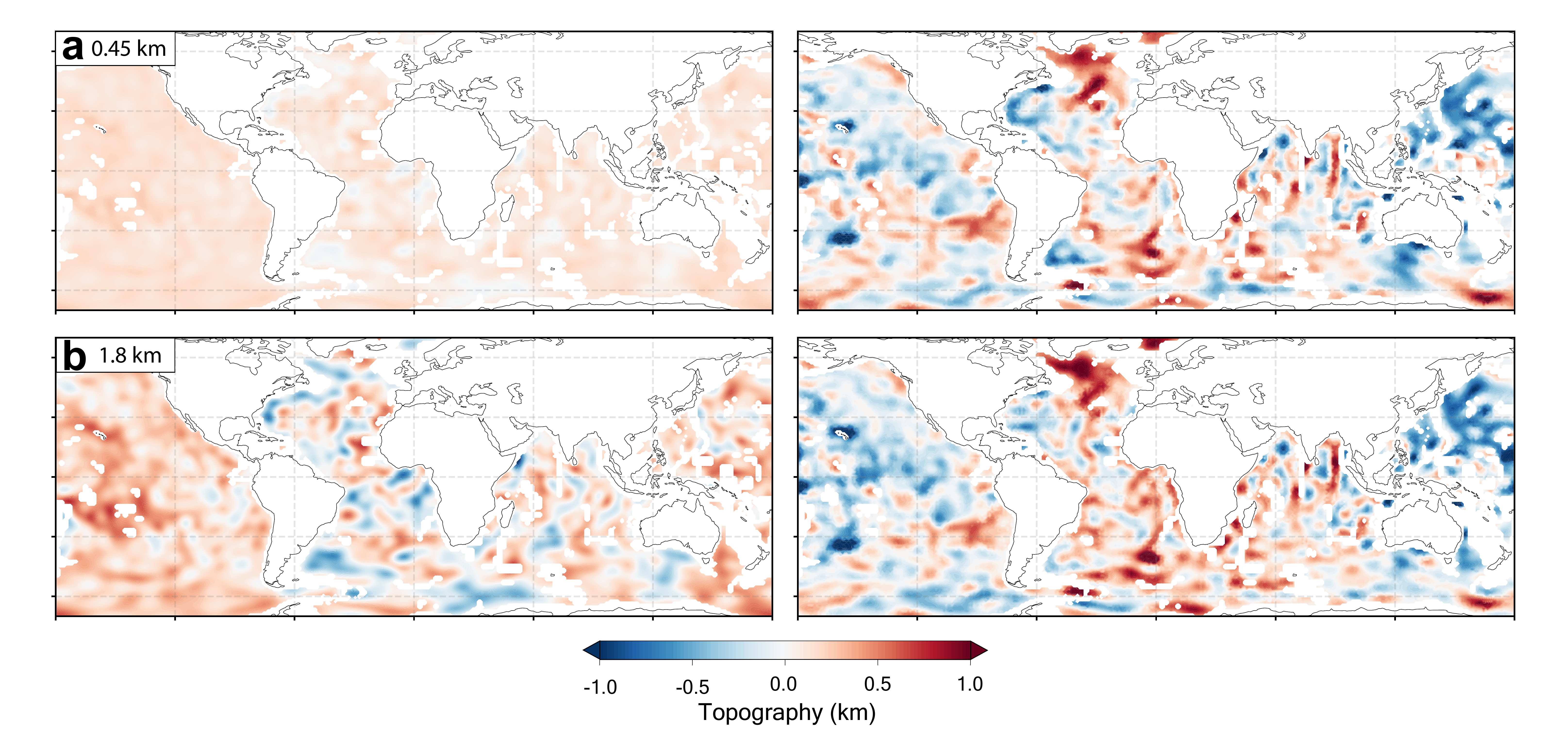}
    \caption{Different scaling of isostatic topography contribution from oceanic mantle lithosphere due to thickness variation (left) and the corresponding mantle dynamic topography (right). a) Half and b) twice the peak-to-trough amplitude of lithospheric residual topography as in Fig.~\ref{fig:fig4}b (0.9~km). The resulting mantle dynamic topography on the right for both cases retains a similar degree-1 dominant pattern as in Fig.~\ref{fig:fig4}c.}
    \label{fig:lithscales}
\end{figure}

\FloatBarrier

\begin{figure}[htbp]
    \centering
    \begin{subfigure}[b]{0.9\textwidth}
        \includegraphics[width=\textwidth]{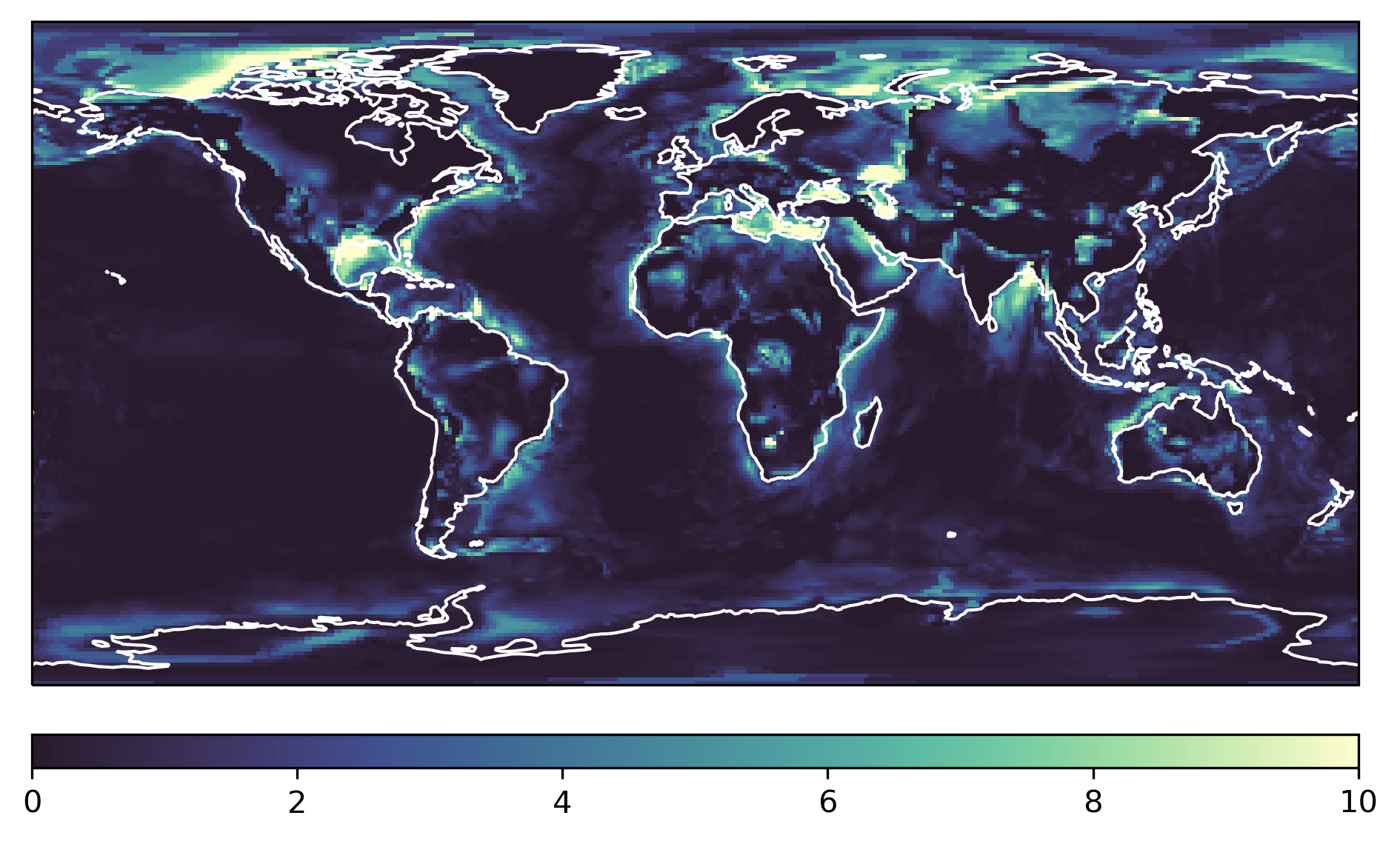}
        \caption{Sediment thickness (km).}
    \end{subfigure}
    \\
    \begin{subfigure}[b]{0.9\textwidth}
        \includegraphics[width=\textwidth]{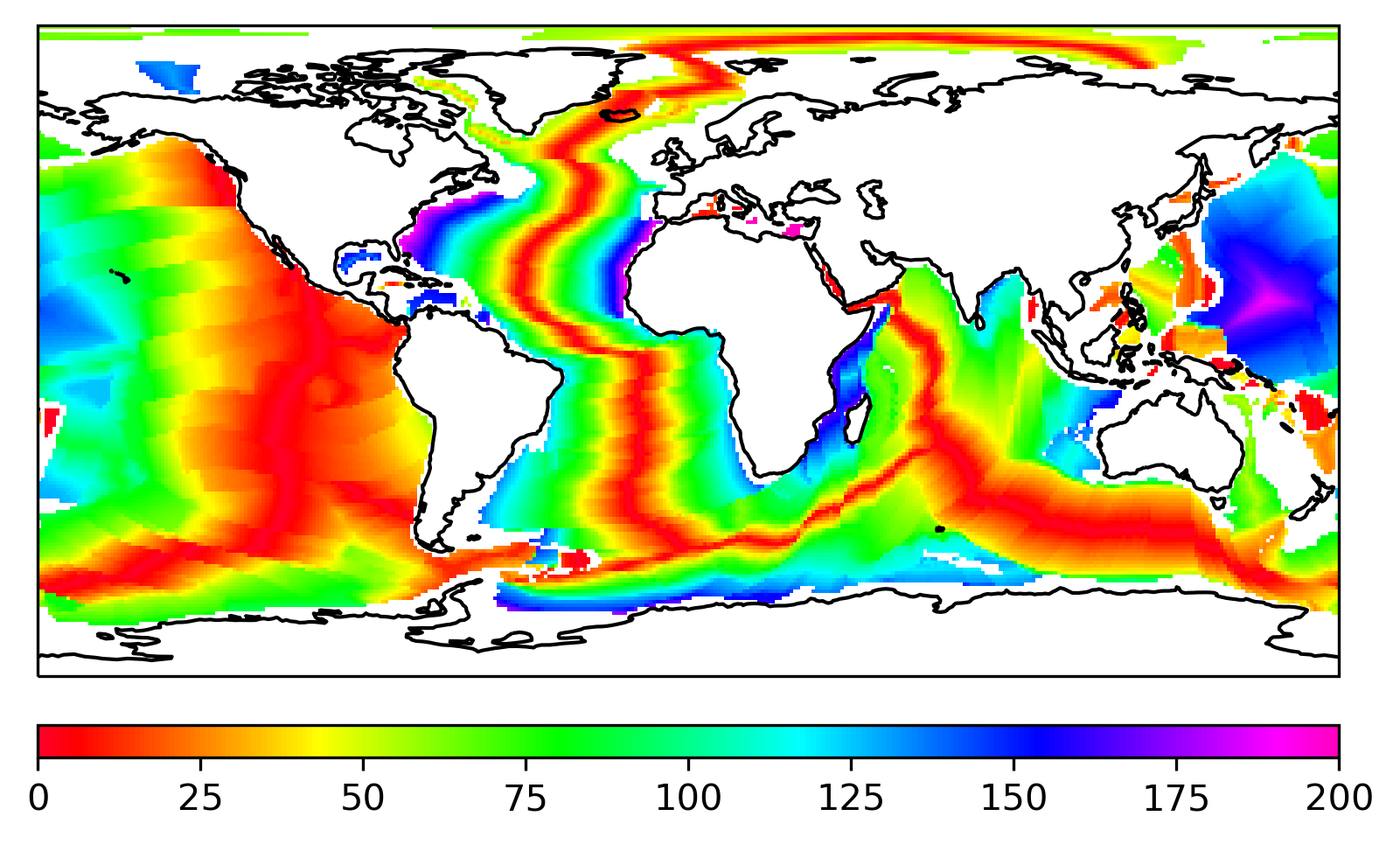}
        \caption{Seafloor age (Ma).}
    \end{subfigure}
    \caption{Comparison between sediment thickness and seafloor age. Although these features have a high degree of correlation in the Atlantic and Indian Oceans, they have little correlation in the Western Pacific and Arctic Oceans.}\label{fig:seafloor}
\end{figure}

\FloatBarrier

\begin{figure}[htbp]
    \centering
    \includegraphics[width=0.7\textwidth]{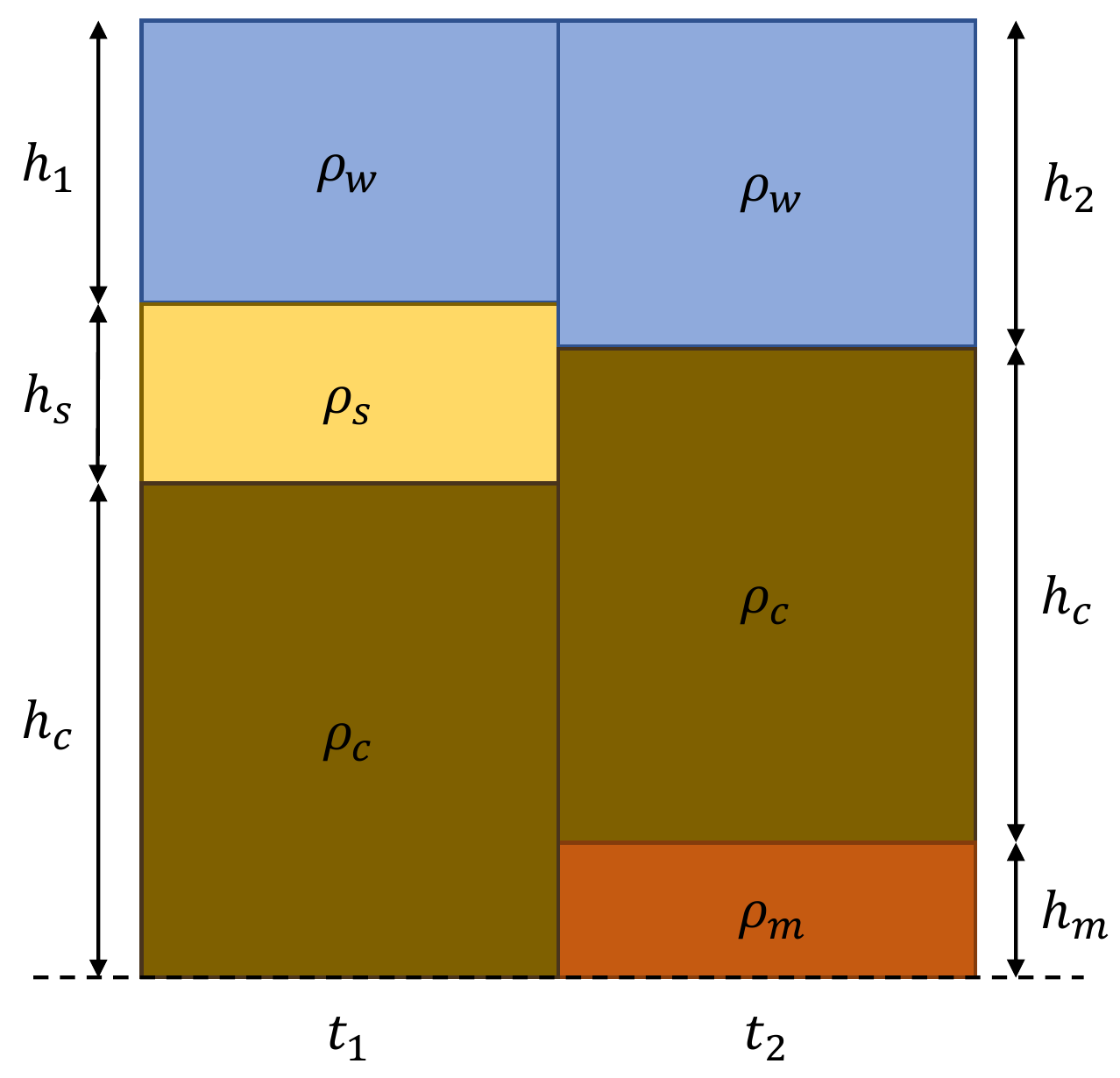}
    \caption{Diagram of isostatic balance equation. Layers include water, sediment, crust, and mantle. Dashed line represents depth of compensation.}
    \label{fig:isostasy}
\end{figure}

\FloatBarrier

\begin{table}[htbp]
    \caption{Performance metrics for robustness study.}
    \label{tab:robust}
    \centering
    \begin{subtable}{0.35\textwidth}
        \centering
        \caption{HS model.}
        \begin{tabular}{lcc}
            \toprule
            Year & RMSE (km) & R\textsuperscript{2} \\
            \midrule
            2020 & 1.051 & 0.214 \\
            2019 & 1.030 & 0.231 \\
            2016 & 1.131 & 0.284 \\
            2013 & 1.154 & 0.259 \\
            2008 & 1.166 & 0.255 \\
            \bottomrule
        \end{tabular}
    \end{subtable}
    \qquad
    \begin{subtable}{0.35\textwidth}
        \centering
        \caption{PSM model.}
        \begin{tabular}{lcc}
            \toprule
            Year & RMSE (km) & R\textsuperscript{2} \\
            \midrule
            2020 & 0.689 & 0.662 \\
            2019 & 0.678 & 0.667 \\
            2016 & 0.729 & 0.703 \\
            2013 & 0.731 & 0.702 \\
            2008 & 0.739 & 0.701 \\
            \bottomrule
        \end{tabular}
    \end{subtable}

    \begin{subtable}{0.35\textwidth}
        \centering
        \caption{GDH1 model.}
        \begin{tabular}{lcc}
            \toprule
            Year & RMSE (km) & R\textsuperscript{2} \\
            \midrule
            2020 & 0.634 & 0.714 \\
            2019 & 0.627 & 0.715 \\
            2016 & 0.675 & 0.745 \\
            2013 & 0.676 & 0.746 \\
            2008 & 0.682 & 0.745 \\
            \bottomrule
        \end{tabular}
    \end{subtable}
    \qquad
    \begin{subtable}{0.35\textwidth}
        \centering
        \caption{H13 model.}
        \begin{tabular}{lcc}
            \toprule
            Year & RMSE (km) & R\textsuperscript{2} \\
            \midrule
            2020 & 0.652 & 0.697 \\
            2019 & 0.645 & 0.699 \\
            2016 & 0.692 & 0.732 \\
            2013 & 0.693 & 0.733 \\
            2008 & 0.699 & 0.733 \\
            \bottomrule
        \end{tabular}
    \end{subtable}

    \begin{subtable}{0.35\textwidth}
        \centering
        \caption{RR model.}
        \begin{tabular}{lcc}
            \toprule
            Year & RMSE (km) & R\textsuperscript{2} \\
            \midrule
            2020 & 0.592 & 0.751 \\
            2019 & 0.590 & 0.748 \\
            2016 & 0.609 & 0.792 \\
            2013 & 0.613 & 0.791 \\
            2008 & 0.613 & 0.794 \\
            \bottomrule
        \end{tabular}
    \end{subtable}
    \qquad
    \begin{subtable}{0.35\textwidth}
        \centering
        \caption{SVR model.}
        \begin{tabular}{lcc}
            \toprule
            Year & RMSE (km) & R\textsuperscript{2} \\
            \midrule
            2020 & 0.564 & 0.773 \\
            2019 & 0.560 & 0.773 \\
            2016 & 0.592 & 0.804 \\
            2013 & 0.601 & 0.799 \\
            2008 & 0.557 & 0.830 \\
            \bottomrule
        \end{tabular}
    \end{subtable}
    
    \begin{subtable}{\textwidth}
        \centering
        \caption{MLP model.}
        \begin{tabular}{lcc}
            \toprule
            Year & RMSE (km) & R\textsuperscript{2} \\
            \midrule
            2020 & 0.476 & 0.839 \\
            2019 & 0.478 & 0.834 \\
            2016 & 0.513 & 0.853 \\
            2013 & 0.497 & 0.863 \\
            2008 & 0.462 & 0.883 \\
            \bottomrule
        \end{tabular}
    \end{subtable}
\end{table}

\FloatBarrier

\begin{table}[htbp]
    \caption{Hyperparameter tuning experiments.}
    \label{tab:hyper}
    \centering
    \begin{subtable}{\textwidth}
        \centering
        \caption{Ridge Regression (RR).}
        \begin{tabular}{ccc}
            \toprule
            Parameter & Parameter Sweep & Optimal Parameter \\
            \midrule
            $\alpha$ & $\left\{0, 10^{-3}, 10^{-2}, 10^{-1}, 10^0, 10^1, 10^2, 10^3\right\}$ & $10^2$ \\
            \bottomrule
        \end{tabular}
    \end{subtable}

    \begin{subtable}{\textwidth}
        \centering
        \caption{Support Vector Regression (SVR).}
        \begin{tabular}{ccc}
            \toprule
            Parameter & Parameter Sweep & Optimal Parameter \\
            \midrule
            kernel & \{linear, poly, rbf, sigmoid\} & rbf \\
            $\gamma$ & \{scale, auto\} & scale \\
            $c_0$ & $\left\{0, 10^{-4}, 10^{-3}, 10^{-2}, 10^{-1}\right\}$ & $10^{-4}$ \\
            $C$ & $\left\{10^{-1}, 10^0, 10^1, 10^2, 10^3\right\}$ & $10^0$ \\
            $\epsilon$ & $\left\{0, 10^{-2}, 10^{-1}, 10^0, 10^1\right\}$ & $10^{-1}$ \\
            \bottomrule            
        \end{tabular}
    \end{subtable}

    \begin{subtable}{\textwidth}
        \centering
        \caption{Multi-Layer Perceptron (MLP).}
        \begin{tabular}{ccc}
            \toprule
            Parameter & Parameter Sweep & Optimal Parameter \\
            \midrule
            activation & \{logistic, tanh, relu\} & tanh \\
            solver & \{lbfgs, sgd, adam\} & adam \\
            $\alpha$ & $\left\{10^{-3}, 10^{-2}, 10^{-1}, 10^0, 10^1\right\}$ & $10^{-1}$ \\
            hidden layers & \{3, 4, 5, 6, 7\} & 6 \\
            hidden size & $\left\{2^7, 2^8, 2^9, 2^{10}, 2^{11}\right\}$ & $2^8$ \\
            learning rate & $\left\{10^{-6}, 10^{-5}, 10^{-4}, 10^{-3}, 10^{-2}\right\}$ & $10^{-3}$ \\
            \bottomrule
        \end{tabular}
    \end{subtable}
\end{table}

\FloatBarrier

\section{Supplementary Information}

All plate models include a sediment correction factor that accounts for the load of sediments on oceanic crust. This sediment correction factor can be calculated as follows.

Extended Data Fig.~\ref{fig:isostasy} illustrates the isostatic balance equation we are trying to solve. Column $t_1$ represents the current state, while column $t_2$ represents a fictional state where sediments have been removed. We set the depth of compensation to the dashed line at the crust-mantle boundary in column $t_1$. Following Airy isostasy, we assume that all densities remain constant and that crustal thickness does not change during this correction.

Our goal is to solve for the uplift of the oceanic crust, $h_m$, after removing the sediments. We do this using Archimedes' Principle, which states that the weight (and therefore mass) of these two columns must be equivalent in order to achieve isostatic balance. We use the fact that oceanic crust cancels out and that $h_1 + h_s = h_2 + h_m$.
\begin{align}
\begin{split}
    \rho_w h_1 + \rho_s h_s + \cancel{\rho_c h_c} &= \rho_w h_2 + \cancel{\rho_c h_c} + \rho_m h_m \\
    \rho_w h_1 + \rho_s h_s &= \rho_w h_2 + \rho_m h_m \\
    \rho_w h_1 + \rho_s h_s &= \rho_w (h_1 + h_s - h_m) + \rho_m h_m \\
    \cancel{\rho_w h_1} + \rho_s h_s &= \cancel{\rho_w h_1} + \rho_w h_s - \rho_w h_m + \rho_m h_m \\
    \rho_m h_m - \rho_w h_m &= \rho_s h_s - \rho_w h_s \\
    h_m (\rho_m - \rho_w) &= \rho_s h_s - \rho_w h_s \\
    h_m &= \frac{\rho_s h_s - \rho_w h_s}{\rho_m - \rho_w}
\end{split}
\end{align}
In our case, sediments are divided into upper, middle, and lower layers. Thus, the equation becomes
\begin{align}
    h_m = \frac{\rho_{us} h_{us} + \rho_{ms} h_{ms} + \rho_{ls} h_{ls} - \rho_w h_s}{\rho_m - \rho_w},
\end{align}
where $h_s = h_{us} + h_{ms} + h_{ls}$.

\end{document}